\documentclass{aa}
\usepackage{graphicx}
\usepackage{longtable}
\usepackage{supertabular}
\def\be{\begin{equation}}
\def\ee{\end{equation}}
\usepackage{natbib}
\usepackage{amssymb}
\begin{document}
\title{X-ray observations of PKS 0745-191 at the virial radius: Are we there yet?}
\author{D. Eckert\inst{1}, S. Molendi\inst{1}, F. Gastaldello\inst{1,2} \& M. Rossetti\inst{1}}
\institute{INAF/IASF-Milano, Via E. Bassini 15, 20133 Milano, Italy\\
\email{eckert@iasf-milano.inaf.it}
\and
University of California at Irvine, 4129, Frederick Reines Hall, Irvine, CA, 92697-4575, USA}
\abstract{}{We reassess the properties of the ICM at large radii in the galaxy cluster PKS 0745-191 in light of recent \emph{Suzaku} measurements.}{We analyze an archival 10.5 ksec \emph{ROSAT}/PSPC observation to extract the surface-brightness profile of PKS 0745-191 and infer the deprojected density profile. We then compare the \emph{ROSAT} surface-brightness profile with the \emph{Suzaku} result. We perform a mass analysis by combining the \emph{ROSAT} density profile and the published temperature profiles from different instruments.}
{We find that the \emph{ROSAT} surface-brightness profile is statistically inconsistent ($7.7\sigma$) with the \emph{Suzaku} result around and beyond the value of $r_{200}$ estimated by \emph{Suzaku}. We argue that, thanks to its large field of view and low background, \emph{ROSAT}/PSPC is to the present day the most sensitive instrument for studying low surface-brightness X-ray emission in the 0.4-2.0 keV band. We also note that the \emph{Suzaku} temperature and mass profiles are at odds with the results from at least two other satellites (\emph{XMM-Newton} and \emph{Swift}).}{The difference in surface brightness between \emph{ROSAT} and \emph{Suzaku} is most likely explained by the existence of additional foreground components at the low Galactic latitude of the source, which were not taken into account in the \emph{Suzaku} background modeling. In light of our mass analysis, we conclude that any estimate of the fraction of the virial radius reached by X-ray measurements is affected by systematic errors of the order of 25\%. As a result, the properties of the ICM at the virial radius are still uncertain, and the \emph{Suzaku} results should be considered with caution.}
\authorrunning{Eckert, D. et al.}
\titlerunning{X-ray observations of PKS 0745-191 at the virial radius: Are we there yet?}

\maketitle
\section{Introduction}

Observing the outskirts of galaxy clusters is important for understanding the formation processes of large-scale structures. For instance, studying the intracluster gas around the virial radius gives an opportunity to measure the transition between the gravitationally-bound gas of clusters and the infalling material from large-scale structures. Tracing the gas out to the virial radius is also important for calibrating the X-ray mass measurements, which are important to cosmology \citep[e.g.,][]{voit}. However, measuring the properties of the intracluster medium in cluster outskirts is difficult because of the low surface brightness of these regions. An instrument with a low internal background is required to measure the source emission at high radii. In the past few years, the \emph{Suzaku} satellite achieved a breakthrough in this domain, performing measurements of the thermodynamical properties of the ICM out to the virial radius \citep{bautz,reip09,hoshino,kawa}.

PKS 0745-191 ($z=0.1028$) is a very luminous \citep[$L_X\sim3\times 10^{45}$ ergs s$^{-1}$ in the 2-10 keV band,][]{arnaud87,allen}, cool-core cluster located in the vicinity of the Galactic plane ($b=+3^\circ$). From a \emph{Suzaku}/XIS observation of the cluster, \citet{george} (hereafter, G09) measured the cluster emission out to $\sim1.5r_{200}$, and determined a value of $r_{200}=1.7$ Mpc (15.2 arcmin) for the virial radius. Surprisingly, the authors noted a flattening of the density and entropy profiles around $r_{200}$, at variance with results from cosmological simulations \citep[e.g.,][]{roncarelli,tozzi}. This and similar results from other authors inspired a large amount of theoretical work \citep[e.g.,][]{lapi,nagai}, invoking several mechanisms (e.g., non-thermal pressure support, gas clumping) to reconcile simulations and observations. An independent confirmation of this result would therefore be very important to our understanding of cluster outskirts.

In this paper, we present the analysis of an archival \emph{ROSAT} Position Sensitive Proportional Counter (PSPC) observation of PKS 0745-191, with the aim of confirming the result of G09. Although the PSPC could not measure temperatures because of its limited bandpass (0.1-2.4 keV), its low instrumental background and large field of view ($\sim$2 square degrees) made it an excellent tool for the study of low surface-brightness regions such as the outer regions of galaxy clusters \citep[see e.g.,][]{vikhlinin99,neumann05}. We also perform a mass analysis using the PSPC density profile and the temperature measurements from various other X-ray satellites, and compare the results with the measurements of G09. The paper is organized as follows. In Sect. \ref{data}, we describe the data analysis procedure. We present our results for the density profile of the cluster in Sect. \ref{results}, and discuss them in Sect. \ref{discussion}.

Throughout the paper, we assume a $\Lambda$CDM cosmology with $\Omega_m=0.3$, $\Omega_\Lambda=0.7$, and $H_0=70$ km s$^{-1}$ Mpc$^{-1}$.

\section{Data analysis}
\label{data}

\subsection{Reduction}

PKS 0745-191 was the target of a pointed \emph{ROSAT}/PSPC observation on October 15, 1993 (observation ID RP800623N00) for a total of 10.5 ksec. We reduced the data using the \emph{ROSAT} Extended Source Analysis Software \citep[ESAS,][]{esas}. To eliminate flaring periods, we extracted a light curve from the raw event files and rejected all time periods where the Master Veto count rate was above 220 counts s$^{-1}$. We then used the ESAS task \textit{\textrm{ao}} to create a model of the scattered solar X-ray background \citep[SSX,][]{solarxrb}, and generated a particle background model using the \textit{\textrm{cast\_part}} executable \citep{partback,plucinsky}. The total model for the non X-ray background (NXB) and the SSX was then inferred.

\begin{figure}
\resizebox{\hsize}{!}{\includegraphics{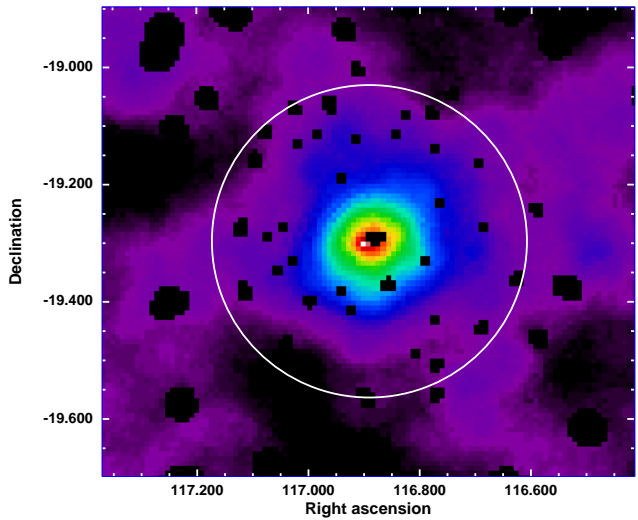}}
\caption{Background-subtracted (NXB+SSX), exposure-corrected, adaptively smoothed \emph{ROSAT}/PSPC image of PKS 0745-191. The black areas indicate the position of masked point sources. The white circle shows the approximate location of $r_{200}$ as estimated by G09.}
\label{image}
\end{figure}

A counts image in the 0.4-2.0 keV band was then extracted from the cleaned event file and the corresponding exposure map was created using the task \textit{\textrm{cast\_exp}}. Point sources were then detected from the image (using the program \textit{\textrm{detect}}) and a point source mask was generated to excise the corresponding regions. As a result, a background-subtracted, exposure-corrected image was created and adaptively smoothed. In Fig. \ref{image}, we show the resulting image together with the position of $r_{200}$ estimated by G09. Point sources have been masked from the image.

\subsection{Density profile}

To measure the density profile of the cluster, we first extracted the surface-brightness (SB) profile of the source following the procedure described in \citet{eckert}. We used the total counts image and the exposure map to extract the mean count rate in concentric annuli out to a radius of 50 arcmin with a bin size of 1 arcmin. The NXB and SSX profiles were also extracted using the same procedure from the background image and subtracted from the total profile.

\begin{figure}
\resizebox{\hsize}{!}{\includegraphics{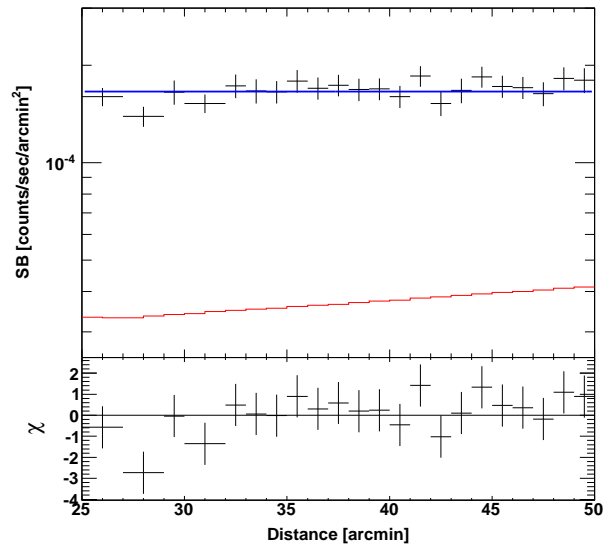}}
\caption{Fit of the SB profile in the 25-50 arcmin radial range by a constant to estimate the cosmic background component (blue). For comparison, the NXB+SSX profile is shown in red. The bottom panel shows the deviations of the data points to the model in units of $\sigma$.}
\label{backfit}
\end{figure}

To subtract the cosmic background component, we fitted the SB profile in the 25-50 arcmin radial range ($r>1.6r_{200}$) with a constant (see Fig. \ref{backfit}). The bins were grouped to ensure a minimum number of 200 counts per bin. The profile in this radial range is accurately-described by a constant ($\chi^2=18.6/21$ d.o.f.). We verified that the background determination does not change significantly when using a different radial range by repeating the same procedure in the radial ranges 25-45, 30-45, 25-40, 30-50, and 35-50 arcmin. In all cases, we find that the measured background value is consistent to within $1\sigma$ with the value extracted from the total 25-50 arcmin range, which indicates that our background determination is stable. The resulting background value was then subtracted from the total profile, and the uncertainties in the background determination were propagated to the background-subtracted profile. In Fig. \ref{backfit}, we show the surface-brightness profile in the 25-50 arcmin range together with the best-fit value for the sky background (blue) and the NXB+SSX profile (red). We can see that at all radii the cosmic component is dominant relative to the NXB and the SSX.

From the background-subtracted profile, we used the procedure of \citet{kriss} to deproject the surface-brightness profile. We assumed the temperature profile of G09, and used XSPEC v12.6.0 and the PSPC effective area to convert the PSPC count rate per unit volume into emissivity using the normalization of the MEKAL model

\begin{equation} \mbox{Norm}=\frac{10^{-14}}{4\pi [d_A(1+z)]^2}\int n_e n_H\, dV,\label{meknorm}\end{equation}

\noindent where $n_e\sim1.2n_H$. We note that in the 0.4-2.0 keV band the conversion factor is insensitive to temperature. For temperatures between 2 and 8 keV, the conversion from PSPC count rate to emissivity changes by at most 4\%. Assuming spherical symmetry and a constant density within each shell, we deduced the density profile.

\subsection{Error estimate}

A crucial point for our analysis is the estimate of statistical uncertainties in the density profile. To compute the error bars in the density profile, we used a Monte Carlo approach to analyzing the \emph{ROSAT} image. Assuming Poisson statistics, we generated $10^8$ realizations of the PSPC counts profile, and performed the procedure described above to obtain $10^8$ realizations of the density profile. The error bars and confidence intervals were then calculated from the distribution of values in each density bin. When a negative background-subtracted profile was found, the density was set to 0. We then defined a 90\% upper limit by the value for which 90\% of the simulations give a result below this value.

\section{Results}
\label{results}

\begin{figure}
\resizebox{\hsize}{!}{\includegraphics{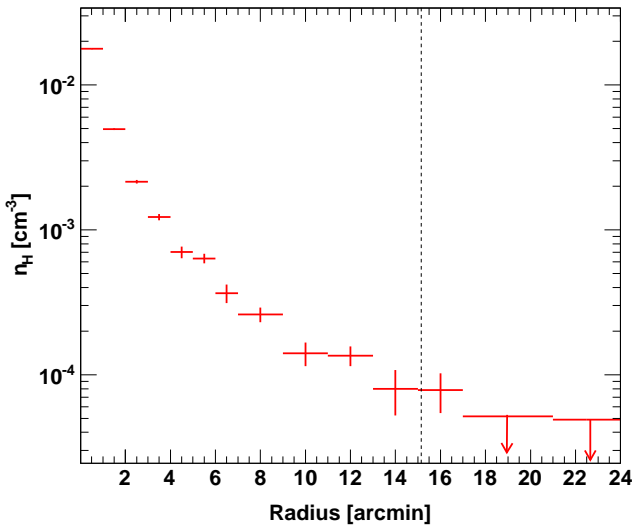}}
\caption{\emph{ROSAT}/PSPC density profile. The error bars and upper limits were estimated from Monte Carlo simulations (see text). The dotted black line shows the approximate location of $r_{200}$ computed by G09.}
\label{density}
\end{figure}

In Fig. \ref{density}, we show the resulting \emph{ROSAT}/PSPC density profile. The dotted black line shows the estimate of $r_{200}$ from G09. Beyond $r=17^{\prime}$, we do not detect any significant cluster emission, and set an upper limit to the density in the $17-25^\prime$ range of $n_{17-25^{\prime}}<4.2\times10^{-5}$ cm$^{-3}$ (90\% confidence level).

\begin{figure}
\resizebox{\hsize}{!}{\includegraphics{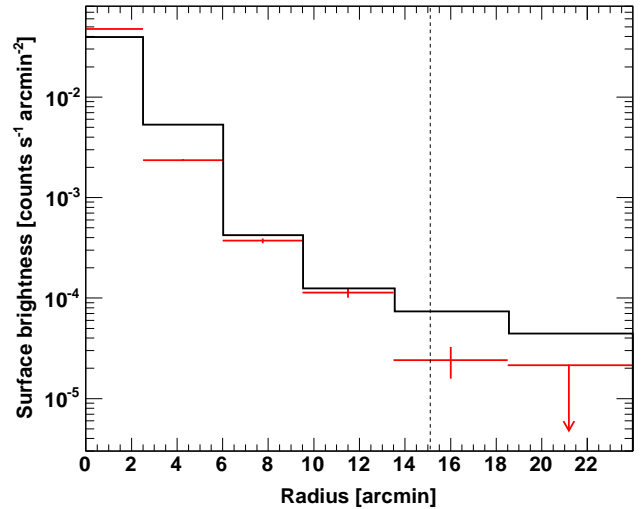}}
\caption{Background-subtracted (cosmic+NXB) \emph{ROSAT}/PSPC surface-brightness profile (red), compared to the \emph{Suzaku} profile from G09 folded through the PSPC response (black).}
\label{projected}
\end{figure}

To compare the \emph{ROSAT} and \emph{Suzaku} profiles, we used a method similar to the one reported in \citet{bowyer}. More specifically, we converted the projected \emph{Suzaku} profile into the MEKAL normalization (Eq. \ref{meknorm}), and folded it through the response of \emph{ROSAT}/PSPC to compute the count rate that should be seen by PSPC. The resulting profile (black), compared to the background-subtracted PSPC profile with the same binning (red), is shown in Fig. \ref{projected}. While in the inner 4 annuli the profiles differ by only $\sim11$\% (the difference in the innermost 2 bins being explained by the broader \emph{Suzaku} PSF), a clear discrepancy between the \emph{ROSAT} and \emph{Suzaku} profiles is found beyond $13.5^{\prime}$. In the 13.5-18.5 arcmin bin, the \emph{Suzaku} profile exceeds the \emph{ROSAT} data point by a factor of three, while in the 18.5-24 arcmin bin our 90\% upper limit lies a factor of 2.1 below the \emph{Suzaku} detection. The profiles at the largest radii are discrepant at more than $7.7\sigma$. This result is stable in terms of the background level. We indeed reach the same conclusion when we use the lowest allowed value for the background instead of the mean value.

\section{Discussion}
\label{discussion}

As shown above, our \emph{ROSAT}/PSPC density profile is statistically inconsistent with the \emph{Suzaku} profile from G09 at high significance. We discuss here several possibilities to reconcile the two results.

\begin{enumerate}
\item
\textit{Cross-calibration:} As found in the case of \emph{ROSAT}/PSPC and \emph{XMM-Newton}/MOS2 \cite[see Appendix A of ][]{eckert}, there could be a cross-calibration discrepancy between PSPC and \emph{Suzaku}/XIS. However, as shown in Fig. \ref{projected} below 13.5 arcmin the projected profiles agree to within 11\%, while at the largest radii the \emph{ROSAT} data points are below the \emph{Suzaku} measurements by a factor of $2.1-3.0$. Thus, although a cross-calibration problem may exist, it certainly cannot explain the observed discrepancy.\\

\item
\textit{Absorption:} Because of its softer bandpass, \emph{ROSAT} is more affected than \emph{Suzaku} by absorption, which is large at the position of PKS 0745-191 ($N_H=4.2\times10^{21}$ cm$^{-2}$). If the Galactic absorption is significantly larger in the outskirts of the cluster ($N_H\gtrsim10^{22}$ cm$^{-2}$) than in the center, the flux observed in the \emph{Suzaku} data can be suppressed in the \emph{ROSAT} band. Variations in the $N_H$ on these scales ($\sim200$ arcmin$^{2}$) should produce significant variations (by a factor of $\sim2.2$) in the CXB intensity in the \emph{ROSAT} band. However, an analysis of the background beyond 25 arcmin from the cluster center did not show any significant variations in the background level. It therefore seems very unlikely that these variations are present only along the line of sight of the cluster.\\

\item
\textit{Spatial coverage:} Because of the smaller field of view (FOV) of \emph{Suzaku}, the G09 result was extracted in a rather small region (see Fig. 1 of G09). If the gas density varied significantly at large radii \citep[as in the case of A1795,][]{bautz}, the regions selected in the \emph{Suzaku} observation might have a higher density, hence would not be representative of the mean density around the virial radius. However, extracting the surface-brightness profile in exactly the same region as G09, we do not find any significant difference from the mean profile. In this specific region, the discrepancy with \emph{Suzaku} is still statistically significant ($5.2\sigma$ in the last two bins).\\

\item
\textit{Background determination:} Because of the very low surface brightness of the cluster emission around $r_{200}$, the background determination is important when measuring the physical properties of the gas in these regions. Thanks to its much wider FOV, \emph{ROSAT} has a clear advantage with respect to \emph{Suzaku}. The large \emph{ROSAT} FOV allows us to perform a local background measurement, while G09 had to rely on background modeling. At the low Galactic latitudes where PKS 0745-191 is located, the foreground emission associated with our galaxy is known to be stronger and more variable than at higher latitudes, making background characterization a more complicated business. Thanks to its higher spatial resolution and larger FOV, \emph{ROSAT} is able to more accurately measure the total sky background emission, and resolve a larger fraction of sources (see Fig. \ref{image}), so the systematic uncertainties affecting the \emph{ROSAT} measurements are smaller than those of \emph{Suzaku}.

An additional background component is probably causing the discrepancy. G09 used an earlier observation of the Lockman hole to estimate the cosmic background component, arguing that, since the total cosmic background level measured by \emph{ROSAT} in the 1-2 keV band at the position of PKS 0745-191 is similar to the one for the Lockman hole, the same model could be used for the sky background in the two observations. However, since the extragalactic background at the low Galactic latitude of PKS 0745-191 is lower because of the larger absorption, another component, most likely Galactic, must be responsible for compensating the decrease in the extragalactic emission. \citet{masui} presented the \emph{Suzaku} spectrum of an empty field located only 3$^\circ$ away from PKS 0745-191. The authors found clear evidence of another sky component at a temperature of $\sim0.8$ keV, with excess emission in the 1-1.2 keV range suggesting that even higher temperature emission exists. This implies that the spectrum is qualitatively unlike empty-field spectra extracted at high Galactic latitude. Since G09 neglected these components, their results are likely to be affected by an incorrect background model, implying that clusters located at low Galactic latitude (or in any region with complicated soft emission, such as the North polar spur) are not ideal targets for the study of cluster outskirts.
\end{enumerate}

In low-SB regions, the key feature for an instrument is the ratio of background level to effective area \citep{ettoriwfxt}. In this respect, \emph{ROSAT}/PSPC is a more sensitive instrument than \emph{Suzaku}/XIS. From the total background level (see Fig. \ref{backfit}) we get a value of $4\times10^{-7}$ counts s$^{-1}$ keV$^{-1}$ arcmin$^{-2}$ cm$^{-2}$ for this quantity (0.4-2.0 keV band), which is almost a factor of two lower than for \emph{Suzaku}/XIS \citep[see Fig. 5 of ][]{mitsuda}.

As stated in G09, the mass parameters derived from the \emph{Suzaku} data appear to be inconsistent with other results. We performed a mass analysis using the \emph{ROSAT} density profile and the temperature profiles from various satellites (\emph{Suzaku}, \emph{XMM}, \emph{BeppoSAX} and \emph{Swift}, see Appendix A). Interestingly, we see that all satellites except \emph{Suzaku} find a value of $r_{200}$ that is larger than 2 Mpc ($\sim25\%$ larger than the value estimated by G09), corresponding to a difference of a factor of 2 in the virial mass. For \emph{XMM} and \emph{Swift} measurements the difference in $r_{200}$ is statistically highly significant, at more than 6 and 5$\sigma$, respectively (see Appendix A for details). 

Since there is no compelling reason to prefer one set of measurements with respect to another, we are forced to conclude that the scale radius in PKS 0745-191 is currently affected by a systematic indetermination of roughly 25\%. It goes almost without saying that any estimate of the fraction of the virial radius reached  by X-ray measurements will have to take this indetermination into account. This is no small issue. Depending on whether we assume the \emph{Suzaku} or the \emph{XMM}/\emph{Swift} virial radius we have that the PSPC surface brightness measurements extend to  $1.03r_{200}$ or $0.8r_{200}$.  Finally, although we cannot prefer one estimate of $r_{200}$ to another, there are arguments favoring  the \emph{XMM}/\emph{Swift} measurement over the \emph{Suzaku} one. First of all, we have two independent measurements that are consistent with one another and inconsistent with the \emph{Suzaku} one. Secondly, the \emph{Swift} and \emph{XMM} profiles are consistent with mean cluster temperature profiles measured with \emph{BeppoSAX} \citep{sabrina}, \emph{XMM} \citep{lm08}, and \emph{Chandra} \citep{vikhlinin06}, and with predictions from simulations \citep[e.g.,][]{roncarelli}. This is not the case for the \emph{Suzaku} profile, which shows a much more rapid and significant decline. This is an important point since, as discussed in the Appendix, the difference in $r_{200}$ follows from the difference in the shape of the radial temperature profiles.

\section{Conclusion}

We have reported our analysis of an archival \emph{ROSAT}/PSPC observation of the galaxy cluster PKS 0745-191. We have found that the surface-brightness profile extracted from PSPC data is statistically inconsistent with the \emph{Suzaku} result from G09 at high significance ($7.7\sigma$). At large radii ($>13.5$ arcmin), the predicted count rate exceeds the PSPC data by a factor of $2.1-3$. This difference is most likely due to a problem in the modeling of the background in the \emph{Suzaku} measurement, which is difficult for a narrow-field instrument at the low Galactic latitude of the source. Thanks to its larger FOV and higher spatial resolution, \emph{ROSAT} is able to measure the background more accurately than \emph{Suzaku}.

We have also shown that the rapidly declining temperature profile measured by \emph{Suzaku} leads to total mass and $r_{200}$ estimates at variance with those derived from \emph{Swift} and \emph{XMM} temperature profiles. In the absence of compelling proof favoring one measurement over another, we conservatively conclude that any estimate of the fraction of the virial radius reached by X-ray measures is affected by a systematic error of about 25\% associated with the indetermination in $r_{200}$. In conclusion, we stress that the current observational knowledge of the ICM properties of PKS 0745-191 at large radii is still uncertain, thus the results presented by G09 should be considered with care.

\acknowledgements{We thank Steve Snowden for providing us and helping us with the \emph{ROSAT} Extended Source Analysis Software, David Buote for the use of his mass analysis software, Daisuke Nagai for useful discussions, and Matt George and Andy Fabian for their constructive comments. DE is supported by the Occhialini post-doc fellowship of IASF Milano. FG and MR are supported by ASI-INAF (I/009/10/0 contract). }
\normalsize

\bibliographystyle{aa}
\bibliography{pks}

\newpage

\appendix

\section{Mass analysis with different datasets}

\begin{figure}
\resizebox{\hsize}{!}{\includegraphics[angle=270]{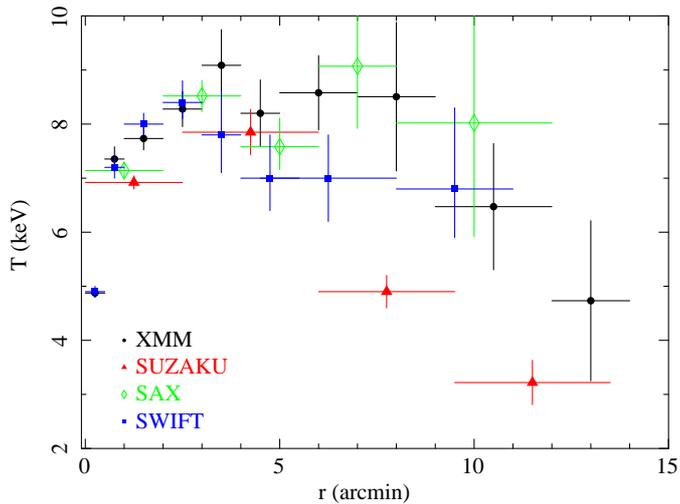}}
\caption{Temperature profiles of PKS 0745-191 measured by \emph{Suzaku} (red triangles, G09), \emph{XMM-Newton} \citep[black circles,][]{xmmcat}, \emph{Swift}/XRT \citep[blue squares,][]{swift} and \emph{BeppoSAX} \citep[green diamonds,][]{sabrina}.}
\label{tprofs}
\end{figure}

We performed a mass analysis adopting method 2 of \citet{ettori10} (and references therein) using the \emph{ROSAT} surface brightness profile (this work). For the temperature profile, we adopted the published temperature profiles of four different instruments: \emph{XMM-Newton} \citep{xmmcat}, \emph{BeppoSAX} \citep{sabrina}, \emph{Swift}/XRT \citep{swift}, and the \emph{Chandra} and \emph{Suzaku} temperature profile of G09 (their Fig.5, i.e. excluding the Chandra data points within 70 kpc and the innermost and outermost Suzaku data points). In Fig. \ref{tprofs}, we
plot the temperature profiles for PKS 0745-191 within the inner 15 arcmin from the center of the cluster, obtained by various satellites with very different instruments (gas proportional counters or CCDs) in very different orbits (low Earth orbits (LEO) with very low particle background such as \emph{BeppoSAX} and \emph{Swift}/XRT, or highly elliptical orbits with high background such as XMM). There is a clear discrepancy between the behavior of the \emph{Suzaku} profile compared to all the other satellites.

In Table \ref{tabmass}, we give the best-fit parameters for the NFW profile estimated using the different datasets. We can see that \emph{XMM-Newton}, \emph{BeppoSAX}, and \emph{Swift} data give results in good agreement, and point towards a value of $r_{200}$ larger than 2 Mpc. 
In contrast, if we use the \emph{Chandra} and \emph{Suzaku} profile of G09 we obtain results in agreement with theirs, $r_{200} = 1.88 \pm 0.03$ Mpc and $M_{200} = (8.32 \pm 0.41) \times 10^{14} M_\odot$, notwithstanding the different methods adopted in the mass analysis, for example the use of the \emph{ROSAT} density profile and the simplifying assumptions of constant density within each shell and estimate of the volume defined by G09. G09 correctly points out the problems of the NFW fits applied to the \emph{Chandra} data because the data do not extend to the NFW scale radius, $r_s$. The mass fits reported in Table \ref{tabmass} all use the \emph{ROSAT} surface brightness profile extending to 1927 kpc and the spectroscopic temperature data out to the radius listed as $r_{xsp}$ for the various satellites. A necessary ingredient for a reliable measurement of the NFW parameters is that the fitted value of $r_s$ lies well within the outer radius of the X-ray data, which is fulfilled by all the adopted datasets. Between \emph{XMM-Newton} and \emph{Suzaku}, the discrepancy is at the $6\sigma$ and $4.8\sigma$ level for $r_{200}$ and $M_{200}$, respectively. The results are at odds at the level of $5.6\sigma$ and $4.6\sigma$ between \emph{Swift} and \emph{Suzaku}, and $1.5\sigma$ and $1.2\sigma$ between \emph{BeppoSAX} and \emph{Suzaku}.

These results clearly highlight the importance of the \emph{Suzaku} temperature profile for the rather low values of $r_{200}$ and $M_{200}$ obtained in G09 compared to those obtained from \emph{XMM-Newton}, \emph{BeppoSAX}, and \emph{Swift} data sets. The discrepancy can be quantified by comparing the value for the slope of the power-law fit to the temperature profile excluding the core as reported by G09, $-0.94 \pm 0.06$, to the mean values of observed temperature profiles beyond $0.2r_{180}$ as reported by \citet{lm08}, $-0.31\pm0.02$, i.e. a $10\sigma$ deviation from the mean. The inconsistency with the milder decline found in hydrodynamical simulations \citep[e.g.,][]{roncarelli} was also noted by G09.

\begin{table*}
\caption{\label{tabmass}Best-fit parameters of the NFW profile to the \emph{ROSAT} density and several published temperature profiles (see Fig. \ref{tprofs}). The parameters are the concentration ($c_{200}$), NFW scale radius ($r_{s}$) in kpc, $r_{200}$ in kpc, and gravitational mass ($M_{200}$) in units of $10^{14}M_{\odot}$. The last parameter, $r_{xsp}$, denotes the maximum radius at which temperature measurements could be performed, in kpc.}
\centering
\begin{tabular}{lcccc}
\hline \hline
\, & \emph{ROSAT}+\emph{XMM} & \emph{ROSAT}+\emph{Suzaku} & \emph{ROSAT}+\emph{SAX} & \emph{ROSAT}+\emph{Swift}\\
\hline
$c_{200}$ & $3.45\pm0.25$ & $9.05\pm0.16$ & $4.89\pm0.89$ & $3.52\pm0.26$\\
$r_s$ & $697\pm77$ & $207\pm2$ & $434\pm137$ & $689\pm76$\\
$r_{200}$ & $2405\pm82$ & $1878\pm33$ & $2124\pm160$ & $2429\pm92$\\
$M_{200}$ & $17.48\pm1.85$ & $8.32\pm0.41$ & $12.03\pm3.04$ & $18.00\pm2.05$\\
$\chi^2$/d.o.f. & 57/23 & 42/20 & 27/18 & 61/19\\
$r_{xsp}$ & 1587 & 2097 & 1360 & 1247\\
\hline
\end{tabular}
\end{table*}

\end{document}